\documentclass[aip,jcp,
numerical,
amsmath,amssymb,
reprint
]{revtex4-1}

\usepackage{graphicx}
\usepackage{xcolor}
\usepackage[version=3]{mhchem}
\usepackage{units}

\begin{document}
\title{Quantifying the error of the core-valence separation approximation}

\author{Michael F. Herbst}
\email{michael.herbst@inria.fr}
\affiliation{CERMICS, {\'E}cole des Ponts ParisTech, 6 \& 8 avenue Blaise Pascal, 77455 Marne-la-Vall{\'e}e, France;
Inria Paris, 75589 Paris Cedex 12, France;
Sorbonne Universit{\'e}e, Institut des sciences du calcul et des donn{\'e}es, ISCD, 75005 Paris, France}

\author{Thomas Fransson}
\email{thomas.fransson@iwr.uni-heidelberg.de}
\affiliation{Interdisciplinary Center for Scientific Computing, Heidelberg University, 69120 Heidelberg, Germany;
Fysikum, Stockholm University, Albanova, 10691 Stockholm, Sweden}

\begin{abstract}
For the calculation of core-excited states probed through X-ray absorption
spectroscopy, the core-valence separation (CVS) scheme has become a vital tool.
This approach allows to target such states with high specificity,
albeit introducing an error.
We report the implementation of a post-processing step
for CVS excitations obtained within the algebraic-diagrammatic construction
scheme for the polarisation propagator (ADC),
which removes this error.
Based on this we provide a detailed analysis of the CVS scheme,
identifying its accuracy
to be dominated by an error balance between two neglected couplings, one between
core and valence single excitations and
one between single and double core excitations.
The selection of the basis set is shown to be vital for a proper description
of both couplings, with tight polarising functions being necessary
for a good balance of errors.
The CVS error is confirmed to be stable across multiple systems,
with an element-specific spread for $K$-edge spectrum calculations
of only about $\unit[\pm0.02]{eV}$.
A systematic lowering of the CVS error by \unit[0.02--0.03]{eV} is
noted when considering excitations to extremely diffuse states, emulating ionisation.
\end{abstract}

\maketitle



\section{Introduction}
For the study of the electronic and atomic structure of materials,
spectroscopy methods in the X-ray regime
have recently seen key advances.%
~\cite{xfelbook2017,stohr,xasxesbook2016, ultrafastrev2018}
By probing the transition of core electrons to (bound) excited states
techniques such as X-ray absorption spectroscopy (XAS)
provide information on the nature of unoccupied
states surrounding specific elements.%
~\cite{stohr, xasxesbook2016, xrayrev2018, xastutorial2006}
The use of core-orbital resonance energies, which are highly
characteristic for the elements, thus provides a local and sensitive tool for
investigating electronic structure.

In order to accurately model the
core-excitation processes including electron relaxation effects is vital.
This requires a theoretical method capable of capturing two aspects:
Firstly, a reduced screening of the probed nuclei
following the removal of a core electron,
leading to a net attraction of the electron density to the core.
Secondly, a smaller repulsive polarisation effect in the valence
region due to interaction with the excited electron.
These counteracting effects have to be properly
accounted for in a theoretical framework,
either by explicitly optimising the excited state
or by introducing (at least) doubly excited configurations.
On top of these modelling difficulties
a practical challenge is that core-excited eigenstates
are embedded in a continuum of valence-excited states,
which prohibits a bottom-up solution of all sought
core-excited for all but the smallest of systems.~\cite{xrayrev2018}
Despite these challenges many theoretical methods
for simulating core-excited spectra have been developed and made available.%
~\cite{tddftrestrict2003,coretddft2010,rewtddft2012a,tpxes1998,
mrcitdxas2019,cumulantXray2019,
ccxspec2,eomccxes2012,cvscc2015,cvseomccsd2019,
cvsadc1980,cvsadc1985,cvsadc2000,cvsadc2014,
nocisxas2018,xrayrev2018}

One approach which has proved to be successful is the
core-valence-separation (CVS) scheme,~\cite{cvsadc1980,cvsadc1985}
in which the weak electrostatic coupling between core- and valence-orbitals
allows certain electron-repulsion integrals (ERI) to be neglected.
Amongst a reduction in the involved matrix dimensions
and thus a decrease in computational cost,
this approximation decouples the valence continuum from core-excited states
allowing to target specifically only the latter kind of excitations.
The CVS approximation has been successfully implemented using a variety of wave
function and density based methods,~\cite{xrayrev2018} including coupled
cluster theory,~\cite{cvscc2015, cvseomccsd2019} density cumulant
theory,~\cite{cumulantXray2019} multireference theory,~\cite{mrcitdxas2019}
time-dependent density functional theory,~\cite{tddftrestrict2003}
and the algebraic-diagrammatic construction
scheme.~\cite{cvsadc1980, cvsadc1985, cvsadc2000, cvsadc2014, opencvsadc2014, ipadc2019}
Note that the application of
a core-valence separation scheme is not unique,~\cite{cvseomccsd2019}
and thus may vary between implementations.

Due to the neglected ERIs or --- as an effect --- the neglected
coupling between core and valence excitation classes,
the CVS approximation introduces an error to the obtained core excitation energies.
Using a range of approaches such as
perturbation theory,~\cite{cvsadc1985}
damped response theory,~\cite{cppcc,adcrixs2017}
real-time propagation approaches,~\cite{tdcc2017}
the Lanczos approach,~\cite{ccxspec2} or by considering the full-space problem,~\cite{cumulantXray2019} 
the CVS error has been reported
to range from close to zero to about $\unit[1]{eV}$,
depending on the employed wave function method, system,
CVS implementation and basis set.
It should be noted, however, that these error values have
been found by considering systems of limited size
and comparatively small basis sets.
Furthermore, applying these methods to larger systems and basis sets may result
in the complication that the core-excited states are no longer
well-separated from the valence continuum,
such that spurious valence-excited states need to be carefully
identified and removed from the region of interest.~\cite{xas4crttddft, ledge2016}
This makes an estimation of the CVS error
for basis sets including polarisation or augmentation rather involved, and
such an assessment of the CVS approximation
for larger basis sets of such types is missing to date.

To overcome this limitation this work presents
a post-processing step based on Rayleigh-Quotient iteration,~\cite{Saad2011}
which amounts to undo the CVS approximation and
refine obtained CVS eigenvectors back to the respective full excitation vectors.
For our study of the CVS error we will apply this \textit{CVS relaxation}
in the context of the algebraic-diagrammatic construction
scheme for the polarisation propagator~(ADC)
using the intermediate state representation (ISR)~\cite{adcisr2004,adcpol2006}.
This family of methods has been demonstrated to give very good
agreement with experiment~\cite{cvsadc2000,cvsadc2014,opencvsadc2014,cvsadc32015}
and it allows a large number of properties in the core region
to be tackled.%
~\cite{cvsadc2000,cvsadc2014,opencvsadc2014,adcvisual2016,adcdynxas2016,adcrixs2017,adcxes2019}
Due to the similarities between ADC and other post-HF methods
such as coupled-cluster theory,
we believe both our CVS relaxation methodology
as well as our results for the CVS error to be applicable
beyond the scope of ADC.

The outline of this paper is as follows:
Section \ref{sec:relxation} summarises the core-valence separation
in the context of ADC and introduces our methodology for CVS relaxation.
Section \ref{sec:comput} lists the computational methodology
for our study of the CVS error on representative compounds
with elements of the second and third period
employing primarily double and triple-zeta basis sets
including (core) polarisation and augmentation.
The obtained results are summarised and discussed in Section \ref{sec:results}.

\section{CVS relaxation in the ADC context}
\label{sec:relxation}
\newcommand{\mat}{\textbf}
\newcommand*{\braket}[2]{\left\langle#1\middle|#2\right\rangle}
The key equation to be solved for the ADC scheme
is the Hermitian eigenvalue problem~\cite{adcreview2015}
\begin{equation}
	\mat{M} \vec{X}_i = \Omega_i \vec{X}_i,
	\qquad \braket{\vec{X}_i}{\vec{X}_j} = \delta_{ij},
	\label{eqn:adc_diagonalisation}
\end{equation}
where one obtains the excitation energies as eigenvalues $\Omega_i$
and the excitation vectors $\vec{X}_i$ as eigenvectors of the so-called
ADC matrix $\mat{M}$. The structure and shape of the ADC matrix $\mat{M}$,
thus the difficulty of
\eqref{eqn:adc_diagonalisation}, depends on the ADC($n$) method under
consideration. Commonly its size prevents a full
diagonalisation,~\cite{adcreview2015} such that iterative methods are employed.
Describing valence excitations at ADC($n$) level
is usually straightforward, since their corresponding excitation energies are
located at the bottom end of the spectrum of $\mat{M}$
making them easily accessible.
In contrast, core-excited states have higher energies and are interior
eigenvalues of $\mat{M}$, such that computing them
by iterative diagonalisation of \eqref{eqn:adc_diagonalisation} is challenging.

\begin{figure*}
	\centering
	\begin{center}
		\includegraphics[width=0.95\textwidth]{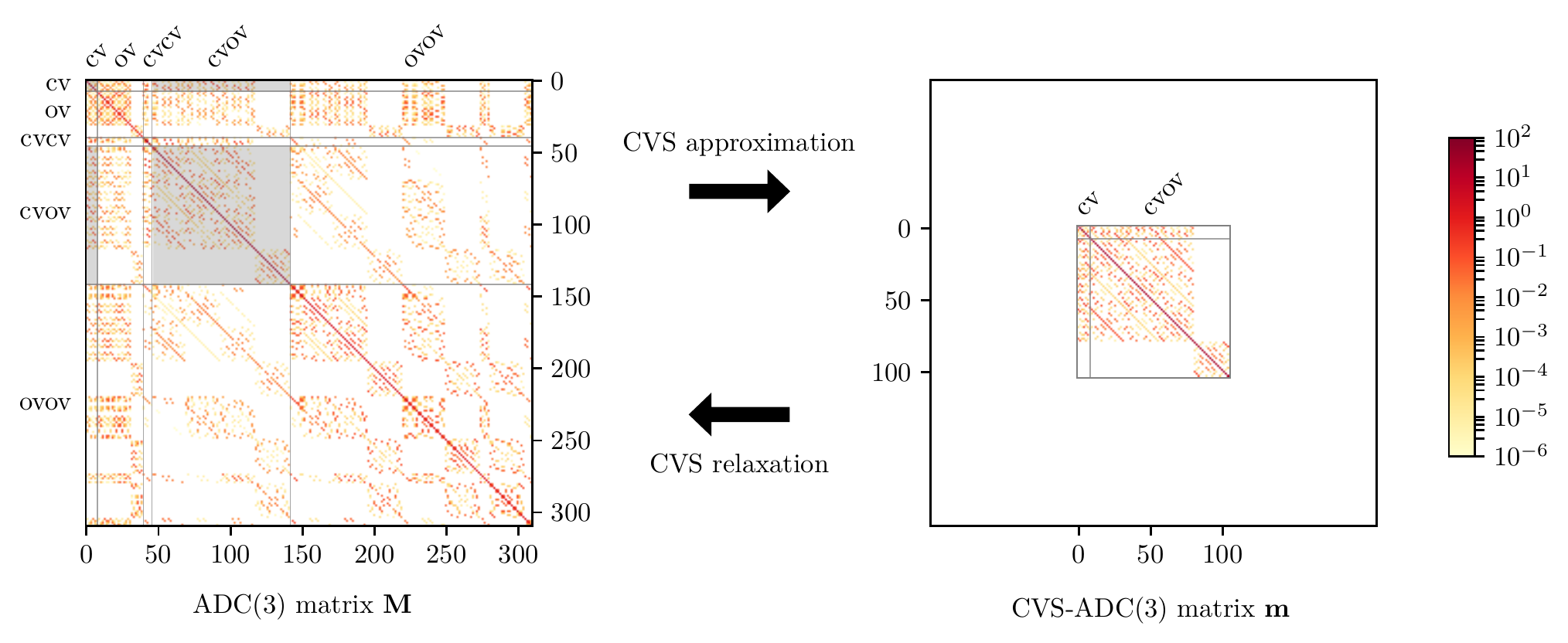}
	\end{center}
	\caption{Full ADC(3) matrix and CVS-ADC(3) matrix of water
		using a STO-3G~\cite{Hehre1969} basis
		obtained using adcc.~\cite{adcc2019}
		Grey lines separate blocks of different
		single- and double-excitation classes
		with ``c'' denoting excitation from a core orbital,
		``o'' from a valence orbital, and ``v'' excitation
		into a virtual orbital.
		Blocks of the full matrix $\mat{M}$,
		which are present in the CVS matrix $\mat{m}$
		as well are marked by a grey background.
	}
	\label{fig:adcmatrix}
\end{figure*}

This picture changes if one considers the CVS approximation.
Following \citet{cvsadc1985} the essence of the CVS scheme is to neglect
the interaction of core and valence orbitals via the Coulomb kernel
by setting a number of electron-repulsion integrals explicitly to zero.
As a result one part of the ADC matrix, which describes
core-excitation processes may be completely decoupled
from the remainder.~\cite{WenzelPhd}
This is summarised in Figure~\ref{fig:adcmatrix}
for the ADC(3) matrix of water.
Denoting by ``o'' a valence-occupied orbital, by ``c'' a core-occupied
orbital and by ``v'' a virtual orbital,
only the parts of the ADC matrix describing the interaction
between cv singly and cvov doubly excited configurations remain
in the CVS-ADC(3) matrix $\mat{m}$.
Notice, that other CVS schemes differ at this point and might, for example,
also include the cvcv doubles excitations in the reduced matrix $\mat{m}$.
Excitation energies and excitation vectors of core-excited states
may now be obtained by solving the smaller eigenproblem
\begin{equation}
	\mat{m} \vec{x}_i = \omega_i \vec{x}_i,
	\qquad \braket{\vec{x}_i}{\vec{x}_j} = \delta_{ij},
	\label{eqn:cvs_diagonalisation}
\end{equation}
where lower-case symbols are used for the equivalent quantities of
\eqref{eqn:adc_diagonalisation}.
Apart from the reduced size an advantage of the CVS approximation
is that the eigenpairs of interest are again located at the
lower end of the spectrum of $\mat{m}$,
such that the same iterative methods to
compute valence-excitations in $\mat{M}$ can now be used
to compute core-excitations in $\mat{m}$.

A consequence of the CVS scheme being an approximation
is that the CVS excitation vector $\vec{x}_i$
is not an eigenvector of the full ADC matrix $\mat{M}$.
The observed validity of the approximation%
~\cite{cvsadc1985,cvscc2015,adcrixs2017,cumulantXray2019}
suggests, however, that it should already be very close.
A good starting point for a refinement procedure
resulting in the full excitation vector $\vec{X}_i$
and corresponding full excitation energy $\Omega_i$,
is thus to start from an already obtained CVS vector $\vec{x}_i$.
This step, the CVS relaxation, thus
effectively undoes the CVS approximation contained in $\vec{x}_i$.

For this purpose we employ a simple scheme based
on Rayleigh-Quotient iteration.~\cite{Saad2011}
applied to one CVS eigenvector at a time.
Starting from the
CVS quantities $\omega^{(0)} = \omega$ and $\vec{x}^{(0)} = \vec{x}$, we
iterate as follows:
\begin{enumerate}
	\item Solve the equation
		\begin{equation}
			\left[\mat{M} - (\omega_i^{(n-1)}+\varepsilon) \,\mat{I} \right]
			\alpha_i^{(n)} \vec{x}_i^{(n)} = \vec{x}_i^{(n-1)}
			\label{eqn:cg}
		\end{equation}
		for $\vec{x}_i^{(n)}$, using a conjugate-gradient
		algorithm preconditioned with the diagonal of the shifted ADC matrix.
		In this $\alpha_i^{(n)}$ is chosen to keep $\vec{x}_i^{(n)}$ normalised
		and $\varepsilon$ is a small positive constant to improve conditioning.
	\item Compute the updated eigenvalue estimate
		\mbox{$\omega_i^{(n)} = \vec{x}_i^{(n)\ T} \mat{M} \, \vec{x}_i^{(n)}$}.
		Converge if residual norm
		\mbox{$\|\mat{M} \,\vec{x}_i^{(n)} - \omega_i^{(n)} \vec{x}_i^{(n)}\|$}
		is below a threshold, else increment $n$ and return to 1.
\end{enumerate}
During the iterations we monitor the overlap
$\braket{\vec{x}^{(n)}}{\vec{x}}$ to ensure
we stay close to the state of interest.
If this value gets too small we restart the algorithm with
based on the iterate of largest overlap and a refined guess for $\omega^{(0)}$.

The discussed relaxation scheme is general
and could be applied to any Hermitian eigenvalue problem,
including alternative formulations of the CVS approximation or
the eigenproblem arising in TDDFT.
Moreover, CVS relaxation in the context of coupled-cluster
could be approached similarly if
equivalent algorithms suitable for non-Hermitian matrices are employed,
such as the generalised minimal residual method~\cite{Saad2011}
for solving \eqref{eqn:cg}.
We remark that we selected this scheme mainly due to its simplicity and generality.
If a black-box application of the CVS relaxation
in production calculations was desired we would expect
further improvements with respect to reliability and performance to be necessary.

%


\section{Computational details}
\label{sec:comput}

Molecular structures have been optimised at the
MP2~\cite{mp2}/cc-pVTZ~\cite{dunning} level of theory,
using the Q-Chem~\cite{qchem4.0} program.
Calculations of core-excited states have been performed
in the adcc python package~\cite{adcc2019,adcc0132}
using Hartree-Fock~(HF) references obtained in pyscf.~\cite{pyscf2018}
For comparing CVS errors,
excitation vectors are first obtained at
either CVS-ADC(1), CVS-ADC(2), CVS-ADC(2)-x or CVS-ADC(3) level of theory
using adcc
and afterwards relaxed to the full ADC level of theory
using a Python implementation of
the CVS relaxation of Section \ref{sec:relxation}.
Absorption cross-sections were computed using the
intermediate state representation (ISR)~\cite{adcisr2004}.
This implementation is available on github~\cite{cvs_github}
and will be integrated into adcc in the future.
For our study we employed the
6-311++G**~\cite{6311Gstarstar} and the Dunning family of basis
sets,~\cite{dunning} including core-polarising~\cite{pCV} and
diffuse~\cite{aug} functions.
Where values are compared to experiment,
relativistic effects were estimated by calculating the
1s MO energies at a HF/cc-pCVTZ level of theory
in both a non-relativistic framework and a scalar relativistic framework
using the second-order Douglas--Kroll--Hess Hamiltonian.%
~\cite{Kroll1974,Hess1986,jansen1989}
These calculations were performed in the Dalton quantum chemical program.%
~\cite{dalton1,dalton2}

\section{Results and discussion}
\label{sec:results}

\subsection{Components of the core-valence separation error}
\label{sec:components}
To leading order, core-excitations are described by the ADC
matrix elements in the cv-cv block of the ADC matrix.
Within the core-valence separation scheme discussed in Section \ref{sec:relxation}
only the interaction of the cv-part of the excitation vector
to the cvov doubly excited configurations is taken into account.
Inside the singles block of the ADC matrix it is thus the coupling
to the ov configurations and inside the doubles block
to both the cvcv and the ovov configurations, which are neglected
and make up the CVS error.
The cv-ovov coupling can be argued to be extremely small
due to both the energetic as well as the spatial separation
of the core and valence orbitals~\cite{cvsadc1980}
and will henceforth be ignored in our discussion.
Similarly all higher-order couplings involving
doubles excitations from different blocks are small and will be ignored.
To leading order we are thus left with the effects
from neglecting the cv-ov and the cv-cvcv coupling within the CVS approximation.
Starting from the CVS result as a reference,
a perturbation theory argument (see Appendix) allows to draw two conclusions:
(1) The cv-ov coupling pushes the energy of the core-excitations \emph{up},
since the ov singles excitations are energetically below the cv,
and (2) the cv-cvcv coupling pushes the energy \emph{down},
since the cvcv doubles have an excitation energy above the cv singles excitations.
The neglected couplings therefore have opposing
effects with respect to the total CVS error.
We note that the present discussion is focused on the $K$-edge
where the cv block only consists of $1s$ core excitations.
For the $L$-edge additional aspects, such as the coupling
between inner core and outer core regions inside the cv block,
can be expected to influence the error balance as well.

For obtaining separate estimates for both sources of the CVS error
we perform two different kinds of CVS relaxations.
First a conventional CVS-ADC calculation followed by a CVS relaxation,
yielding the \emph{total CVS error}.
Second a calculation, where only the singles block
of a particular ADC level of theory is used
in both the CVS-ADC calculation as well as the relaxation.
We will refer to the error observed in this case
as the \emph{singles-block CVS error}.
By construction the results of the singles-only CVS calculation
differ from their relaxed counterparts only by neglecting the cv-ov coupling.
To leading order the singles-block error is therefore equal to the part of
the total CVS error originating from the neglected cv-ov coupling.
This implies that the difference between total and singles-block CVS error
provides a leading-order estimate to the error
from neglecting the cv-cvcv coupling in the CVS treatment.

\begin{figure}
\begin{center}
\includegraphics[width=0.95\columnwidth]{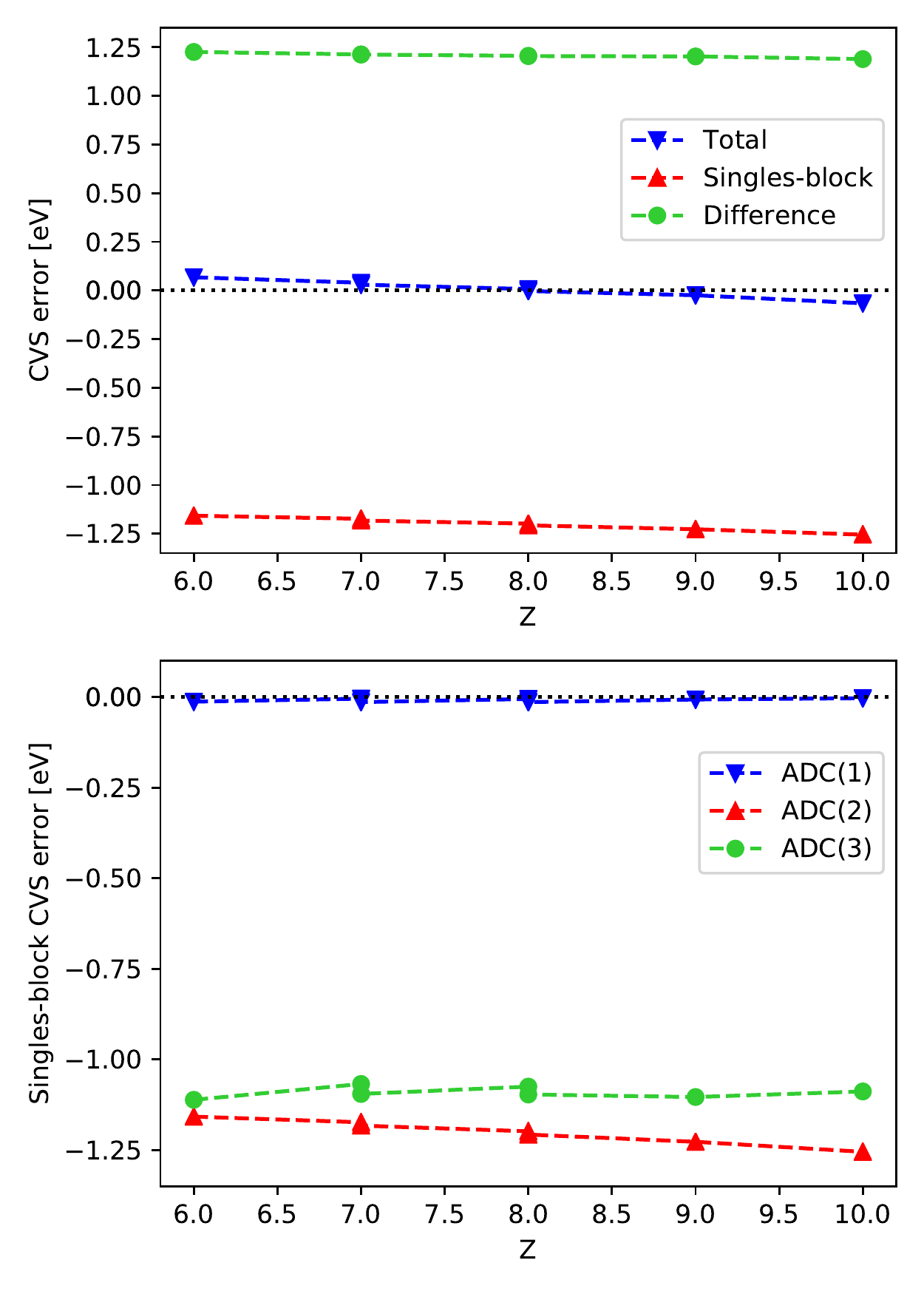}
\caption{The CVS error of the 10-electron series, using an aug-cc-pCVTZ basis
set on C, N, O, F, and Ne, and cc-pVTZ on H. Top: CVS error broken down to
full error, singles-block error, and error difference, obtained using
ADC(2). Bottom: Singles-block CVS error of the ADC hierarchy.}
\label{fig:effects}
\end{center}
\end{figure}
To illustrate the magnitude of these effects,
Figure~\ref{fig:effects} reports the CVS errors
of the 10-electron series, using
an exhaustive aug-cc-pCVTZ
basis on the probed atom and cc-pVTZ on the hydrogens.
Errors have been obtained for one state for each molecule,
save \ce{NH3} and \ce{H2O}, for which two states were included.
The top panel provides an overview of the
total CVS error for ADC(2),
the singles-block CVS error
as well as the difference between these.
As expected from our discussion,
the values of the singles-block CVS error are negative,
meaning that the introduction of the cv-ov coupling
during relaxation raises the energy of the core excitation.
In turn, the difference as a measure for the cv-cvcv coupling has the opposite sign.
We note that in absolute values both errors
are of similar size, amounting to
approximately \unit[1.2]{eV}.
Tests using a core-polarised quadruple-zeta basis set
also yields a small total CVS error.
As will be detailed in Section \ref{sec:basis}
this is not an artefact, much rather it suggests
that a balanced description of both terms
is key to obtain a small CVS error.
In the bottom panel of Figure~\ref{fig:effects}
the singles-block errors for ADC(1), ADC(2) and ADC(3) are reported.
For ADC(1) the error is small, while ADC(2) and ADC(3)
have single-block errors of similar size, being larger for ADC(2).
This is not surprising, since the off-diagonal elements for the ADC(1)
matrix are only made up of the Coulomb and exchange integrals,
which are especially small for the cv-ov coupling block.

\subsection{CVS error behaviour in the ADC($n$) hierarchy}
\label{sec:method}

\begin{figure}
\begin{center}
\includegraphics[width=0.95\columnwidth]{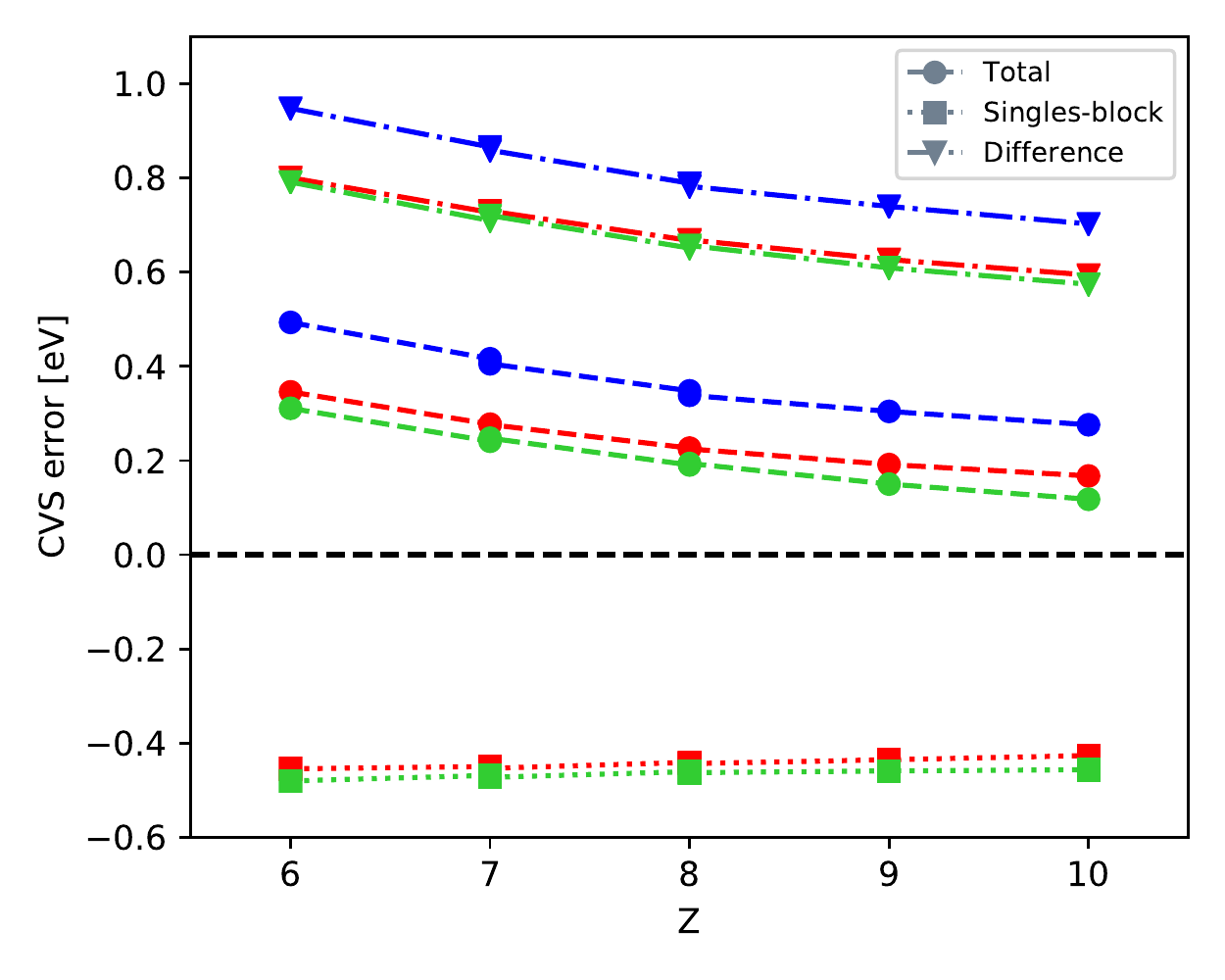}
\caption{Components of the CVS error of ADC(2) (blue), ADC(2)-x (red),
and ADC(3) (green), as a function of atomic number Z.
Results obtained for the 10-electron series, using the 6-311++G** basis set.}
\label{fig:adc_level}
\end{center}
\end{figure}
Figure~\ref{fig:adc_level} summarises the CVS errors
(full, singles-block, and difference)
calculated using the 6-311++G** basis set.
We selected this basis set because it has been shown
to give good agreement with experiment at a relatively low computational cost,
making it important for production calculations.%
~\cite{cvsadc32015,cvsadc2014,eomcc2015,cvseomccsd2019}
It should be noted that all observed trends could be reproduced
with a \mbox{cc-pCVTZ} basis.

With respect to the CVS error
we observe the trend ADC(2) $>$ ADC(2)-x $>$ ADC(3)
for the total error as well as all components.
Most notably, the increased order in perturbation theory
in the doubles block going from ADC(2) to ADC(2)-x
causes a clear drop of total CVS error towards the value observed with ADC(3).
In this basis the singles-block error
turned out to be almost agnostic to the ADC variant
(the singles-block CVS error is identical for ADC(2) and ADC(2)-x, by construction),
such that the difference in errors
shows a similar behaviour as the total error.
Recalling the error difference to be a measure for the cv-cvcv coupling
we attribute the observed trend
to a compensation of the neglected cv-cvcv coupling
by the higher-order treatment
of cvov excitation in
CVS-ADC(2)-x and beyond.
The error in the description of core-relaxation effects,
to which the cv-cvcv couplings contribute,
can thus be expected to be smaller at the CVS-ADC(2)-x and CVS-ADC(3) level
than at CVS-ADC(2). In terms of absorption cross-sections,
the CVS results typically underestimate the intensities obtained in a full treatment.
For the systems presented in
Figure~\ref{fig:adc_level}, for example,
we report discrepancies from \unit[-6]{\%} to \unit[+2]{\%},
with an overestimation observed only for neon.
The trend of a lowering of intensities (with the exception of neon)
is observed for all our calculations, with discrepancies typically being lower
for larger basis sets and for third row elements.
This suggests that the behaviour is general, and our discussion of the CVS error
thus focuses on excitation energies.

\begin{figure}
\begin{center}
\includegraphics[width=0.95\columnwidth]{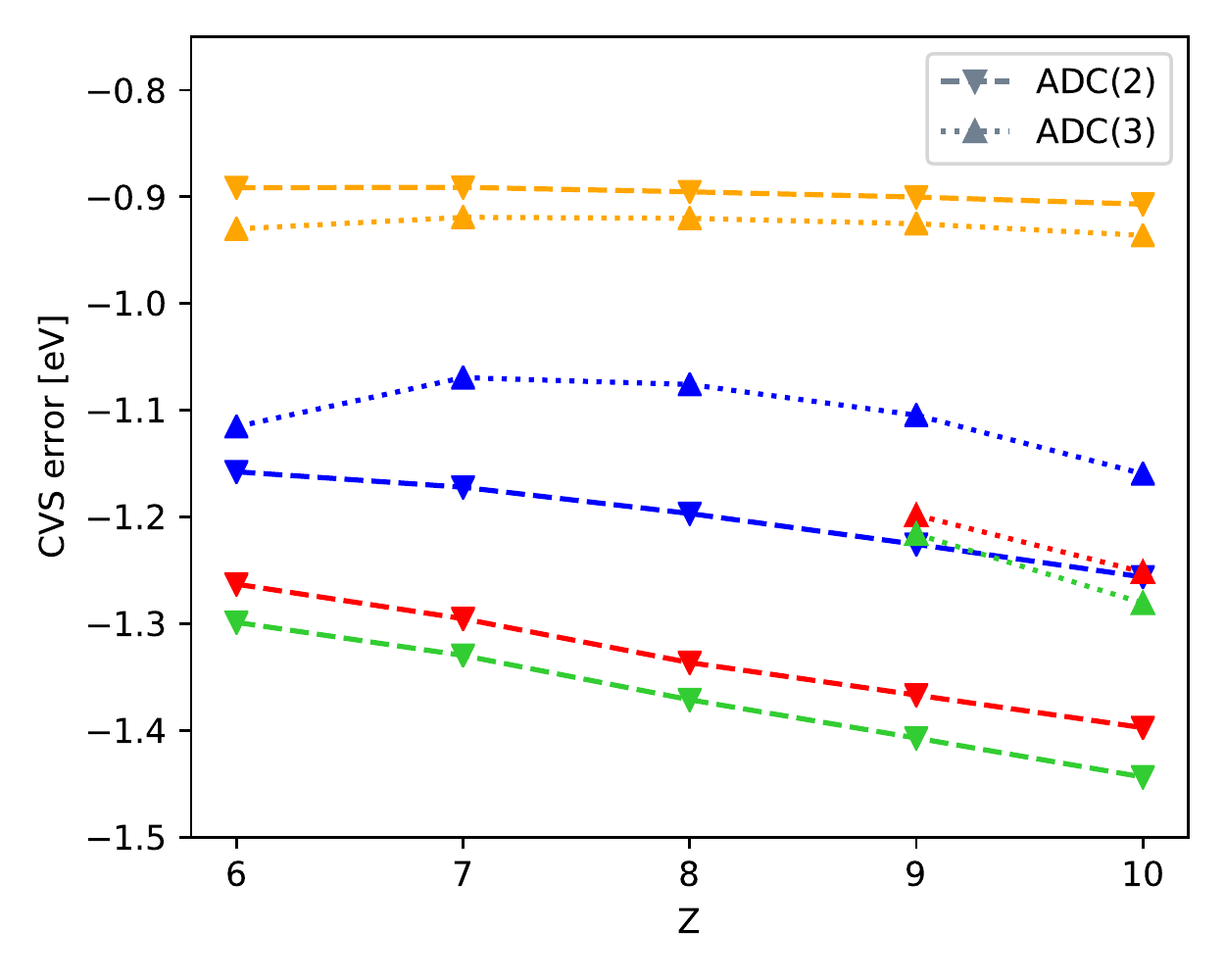}
\caption{Singles-block CVS error of ADC(2) and ADC(3), as obtained using 
cc-pCVDZ (orange), cc-pCVTZ (blue), and cc-pCVQZ (red) basis sets, as well as for a CBS
estimate (green). Results obtained for the 10-electron series.}
\label{fig:cbs}
\end{center}
\end{figure}

For the singles-block CVS error of ADC(2) and ADC(3), we performed a more exhaustive
study using the cc-pCV$n$Z ($n =$ D, T, Q) series
of core-polarised Dunning basis sets,
illustrated in Figure~\ref{fig:cbs}.
Complete basis set extrapolations are included for the systems,
where our CVS relaxation procedure managed to converge
the cc-pCVQZ states.
In general, the size of the singles-block CVS error
increases with the size of the basis.
Over the range of considered elements it is relatively constant
if a double-zeta basis is used.
For larger bases the discrepancy increases with $Z$
for ADC(2), while trends are less clear for ADC(3).
Between ADC(2) and ADC(3) the magnitude of the errors
are similar for cc-pCVDZ, with ADC(3) resulting in slightly larger values.
In the larger basis sets, however, the ordering and relative sizes are
markedly different --- here the ADC(3) error is significantly smaller,
especially using cc-pCVQZ.
The error component due to the neglected cv-ov coupling
is thus smaller in CVS-ADC(3) compared to lower-order
\mbox{CVS-ADC} variants.
Similar to what was observed for the cv-cvcv error
in contrast to the order of perturbation theory in the doubles block,
this suggests that a treatment of the singles-block at higher order
is favourable to recover some effects of the neglected cv-ov interaction,
In our subsequent analysis we will focus on the CVS error in ADC(2),
where the results presented here have identified
both CVS error components to be largest.

\subsection{Basis set effects}
\label{sec:basis}

\begin{figure*}
\begin{center}
\includegraphics[width=0.85\textwidth]{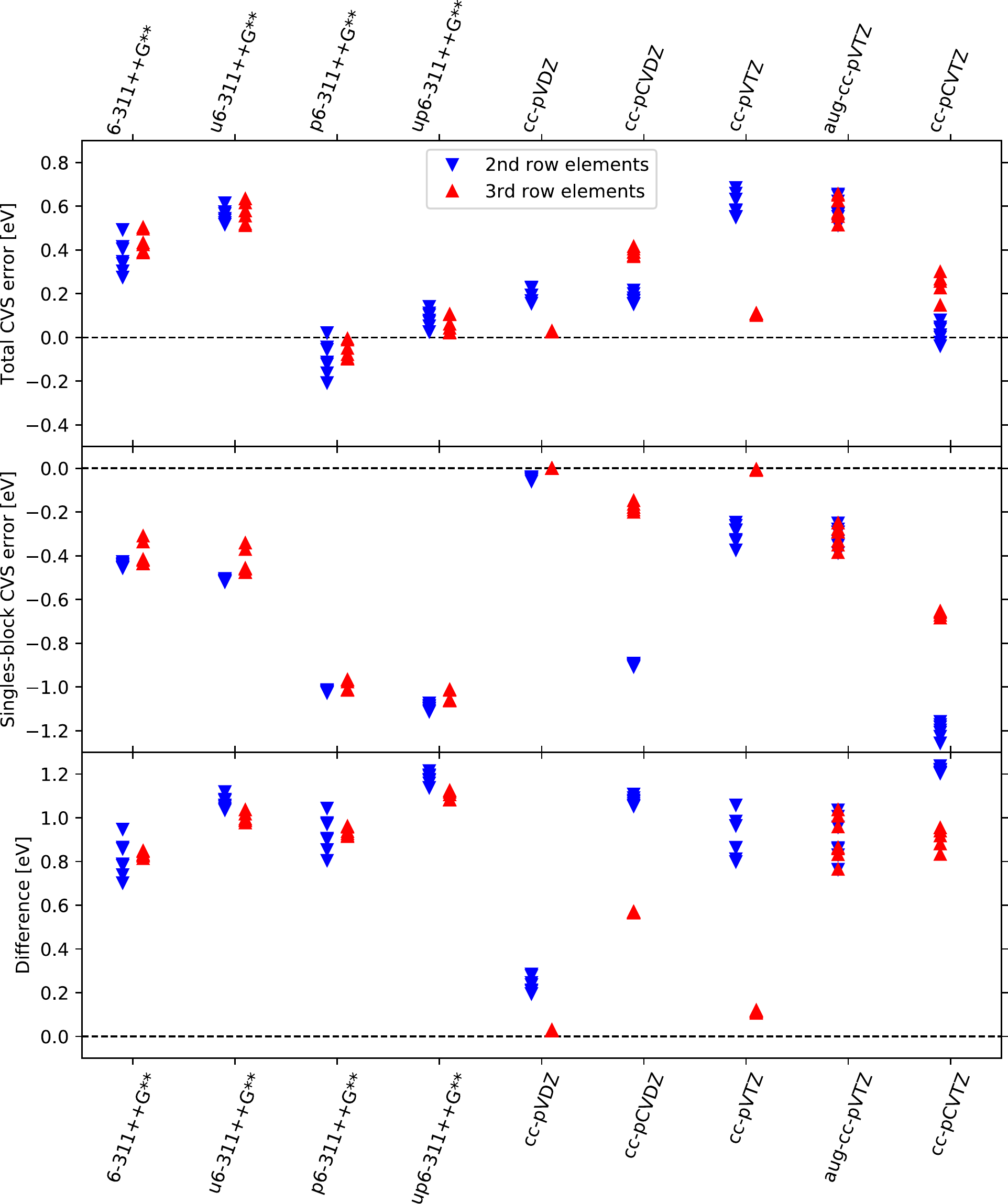}
\caption{The components of the ADC(2) CVS error, as obtained for the 10- and 18-electron series using a number of different basis sets. Note that all panels are to scale.}
\label{fig:basis}
\end{center}
\end{figure*}
Suitable basis sets for core-excitations and core-ionisations
are well-investigated,
with some tendency to favour IGLO,~\cite{iglo1982,xraybasis2018}
basis sets formed by inclusion of functions
from the next element,~\cite{scfipbasis2018}
or amending standard basis sets
by core-polarising functions.~\cite{xraybasis2018,fluoroethenes2013,cvseomccsd2019}
For our investigation of the CVS error we have
focused on 6-311++G** and basis sets from the Dunning family. The former is commonly used for X-ray spectroscopies, and the latter
is known to converge rigorously for correlated methods
and offer a standardised way to include core-polarising and diffuse functions.%
~\cite{Hill2013}
Motivated by previous results with modified 6-311++G** basis sets
yielding good agreement with experiment,~\cite{xestddft2014, xraybasis2018, adcxes2019, xpsbasis2020}
we also included three modifications on 6-311++G**,
namely (1) \mbox{u6-311++G**},
where the $1s$ basis functions have been decontracted,
(2) \mbox{p6-311++G**},
where on top of 6-311++G** we added another six $p$-functions
with exponents taken from the $1s$ functions,
and (3) \mbox{up6-311++G**},
formed by both decontraction and addition of six $p$-functions.%
\footnote{
	Files defining the modified basis sets are available
	from Reference~\onlinecite{cvs_github}.
}
These allow to probe the flexibility required in the core region
for obtaining a low CVS error.
Our results for
the 10-electron (18-electron) series of the second-row (third-row) elements
at the ADC(2) level
are shown in Figure~\ref{fig:basis}.
The trends in the total CVS error~(top panel)
can be best explained by considering them as the sum of the
trends in the singles-block CVS error~(middle panel)
and the difference error~(bottom panel).
Since these latter measures are directly related to the cv-ov and cv-cvcv couplings,
it is in fact the balance between the ability of a basis set to describe
the cv and ov singles and cvcv doubles excitations,
which plays the key role.
This in turn depends on the flexibility of the basis set
to describe both core and valence region in a well-adjusted manner,
such that relaxation and correlation effects in the core
as well as the valence region can be described.

Since most basis sets focus on a description of the chemically most interesting
valence region,
for small basis sets cv- and ov-excitations are not described equally well.
As basis sets get larger or as extra functions in the core region  are
added (e.g.~by core polarisation),
the description of the core orbitals catches up,
leading to a drop in cv-excitation energies relative to the ov-excitation energies.
As a result the gap between them decreases and the cv-ov coupling
and the singles-block CVS error increases.
In our results this can be observed
in the change in singles-block CVS error
between 6-311++G** and p6-6-311++G**,
where the only change to the basis is the addition of polarising $p$-functions
in the core region.
Other examples are the change between cc-pVDZ and cc-pCVDZ,
as well as cc-pVTZ and cc-pCVTZ.

Along similar lines one would expect an improved description
of the core region to increase the cv-cvcv coupling,
since the cvcv doubles excitations are twice as effected by the drop
in core orbital energies as the cv singles excitation.
In this aspect the trends in our results
for the difference between full and single-block CVS error
are not so clear.
Still, the largest error differences in our study
are observed for cc-pCVTZ and up6-311++G**,
i.e.~the basis sets with largest flexibility in the core region we consider,
which suggests that basis sets providing a fuller description
of the core region will indeed have a larger error difference.
Additionally, most other basis sets also yield difference values
in reasonable agreement with cc-pCVTZ and up6-311++G**,
the main outliers being cc-pVDZ for all elements,
and cc-pCVDZ and cc-pVTZ for the third row elements.
Surprising is the significant shift in CVS error difference when
adding diffuse functions to cc-pVTZ for the third row elements, with average
error changing from about \unit[0.1]{eV} to \unit[1]{eV}.
One would not expect the cv-cvcv coupling to be particularly affected
by the addition of diffuse functions.

With respect to the full CVS error,
both neglected couplings contribute with opposite sign, thus
giving rise to an even more unsystematic behaviour.
The smallest observed errors result if a basis describes cv-ov and cv-cvcv coupling
either equally good (e.g.~cc-pCVTZ for the second row)
or equally bad (e.g.~cc-pVDZ for the third row).
It is thus the basis sets with large, but similar, magnitudes
in the singles-block error and the error difference,
which offer a description of the core region,
where core-relaxation is accounted for as best as possible,
but still a small total CVS error results in a CVS scheme.
According to this metric the best basis for the second-row elements in our study
is cc-pCVTZ, whereas for the third-row elements it is up6-311++G**.
For \mbox{cc-pCVTZ} the total CVS errors values are distributed as $\unit[0.02\pm0.04]{eV}$
for the second row and $\unit[0.24\pm0.06]{eV}$ for the third row,
and for up6-311++G** $\unit[0.08\pm0.04]{eV}$ and $\unit[0.06\pm0.04]{eV}$,
respectively.
If smaller basis sets are desired, cc-pCVDZ is reasonable for the second row,
but again p6-311++G** is the better choice for the third row.
We note
that the poor performance for cc-pCVDZ and cc-pCVTZ for the third-row elements
is not surprising, since the core-polarising functions
are the result from minimising the amount of core-core \emph{and}
core-valence correlation combined,~\cite{Hill2013,Jensen2013}
thus putting more emphasis on the outer core region.
Our study, however, probes the inner core region with $1s$ core excitations,
making the standard core-polarised Dunning basis sets less applicable
for the third-row elements.

Summarising the trends obtained over the basis sets in Figure~\ref{fig:basis},
we find that properly describing the cv-ov coupling is more challenging than the cv-cvcv.
Additional $p$-functions or the core-polarised Dunning basis sets
are clearly required to describe the core and valence region
on a similar level.
Considering specifically the 6-311++G** basis set and modifications thereof,
we note that the addition of tight $s$-functions
only amounts to improve the description of the cv-cvcv coupling a little
(raising the error difference),
but it is the addition of the tight $p$-functions
which drastically improves the ov-cv coupling and thus leads to a drop
in total CVS error.
We remark that these conclusions might be different for core spectroscopies
probing $L$-levels, due to stronger coupling between $2s$ and $2p$,
as well as the presence of a deeper lying $1s$
--- indeed, a recent publication on EA- and IP-ADC reported significantly
larger CVS errors for ionisation of $2s$ than of either $1s$ or $2p$.~\cite{ipadc2019}

From our results we expect CVS schemes
which include the cv-cvcv coupling
to lead to a negative CVS error,
i.e. to an overestimation of the excitation energy for $1s$-core excitations.
This is especially the case for the larger and more adequate basis sets,
where they will not benefit from the error cancellation obtained from neglecting both
the cv-ov and the cv-cvcv coupling.
For smaller basis sets, the magnitudes of both cv-ov and cv-cvcv couplings
are generally smaller,
such that we believe this aspect to be overlooked in previous
analysis comparing CVS schemes.

\subsection{The CVS error of different compounds and states}
\label{sec:compounds}

\begin{figure*}
\begin{center}
\includegraphics[width=0.99\textwidth]{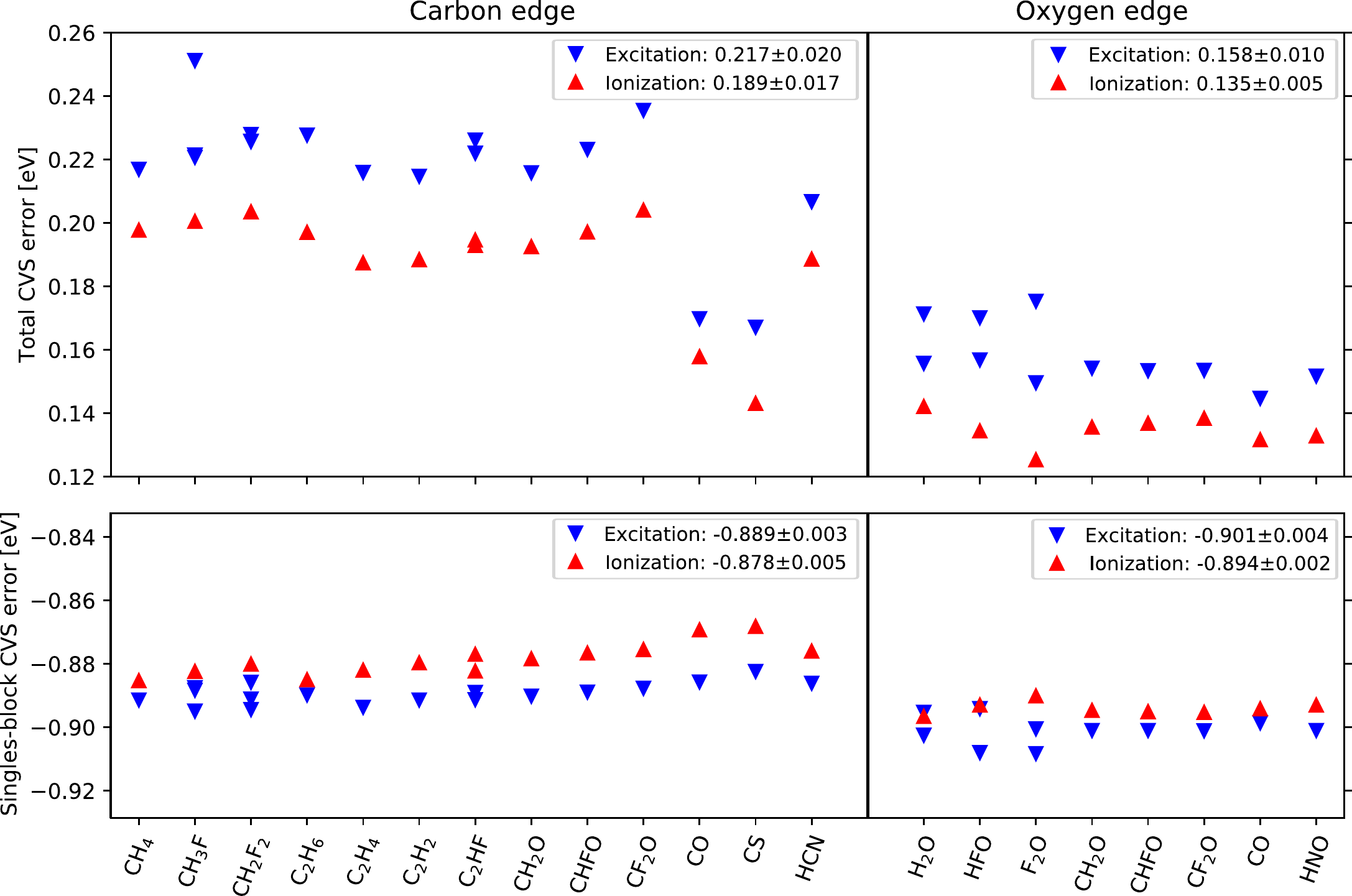}
\caption{The CVS error of ADC(2) for a number of compounds, as obtained using a
cc-pCVDZ (with additional diffuse s-function) basis set for the probed element, and
cc-pVDZ for all other elements. Lower panel reports singles-block CVS error,
using the same energy scale but different error boundaries. Upper
right corner of each panel and edge reports the average error and standard
deviation thereof.}
\label{fig:compounds}
\end{center}
\end{figure*}
We now consider the element-specific spreads of the CVS error.
In Figure~\ref{fig:compounds} we summarise the performance of the CVS approximation
for the carbon and oxygen edge of a number of different compounds.
These results have been obtained using the cc-pCVDZ basis set with an additional
diffuse $s$-function (exponent $10^{-11}$) for the probed element,
and cc-pVDZ for all others.
Intense transitions to bound states as well as ionisations are both
included in our analysis
with the latter being modelled by investigating transitions to the additional diffuse
function.~\cite{diffuseip1999,cvscc2015}
Trends observed in the cc-pCVDZ basis set, already confirmed to yield a
reasonable description of both CVS error sources,
could be reproduced with a up6-311++G** basis set for selected cases.
In agreement with our observation in Section~\ref{sec:method}
intensities for the presented compounds are underestimated
by 1--4 \% when using the CVS approximation.

Over the range of considered compounds we note systematically lower full CVS errors
for ionisation of about \unit[0.02-0.03]{eV}.
For all compounds the singles-block CVS error is of similar size,
with bound excitations being about \unit[0.01]{eV} lower in energy.
This minor difference in errors for both types of excitation processes
is expected, since both are dominated by the electronic structure of the core region
and deviating aspect is the final state
--- once a delocalised $s$-function (ionisation)
and once a localised valence orbitals.
This also explains the smaller total CVS error observed for ionisations,
since the cv-cvcv coupling is small between
a delocalised cv-transition and cvcv-doubles
excitations involving localised virtuals.
Similarly we do not expect the CVS error for ionisation processes
to change significantly if more sophisticated discretisation schemes
for the continuum are employed. Adding additional diffuse functions,
for example, resulted in a very quick convergence of the CVS error
in our experiments.

Notice that our treatment of ionisations can be considered
as a limiting case of transitions to diffuse Rydberg states.
As such the \unit[0.02--0.03]{eV} difference in CVS error is
the mean upper limit of the difference in CVS error between core-valence
and core-Rydberg transitions.
This is confirmed by rough tests on the core-Rydberg CVS error (not shown) yielding
error values between that of excitations to valence states and ionisations.
Overall this difference is negligible,
being one order of magnitude smaller than the CVS error,
which in turn is one order of magnitude smaller
than the error with respect to experiment.
We therefore do not expect this to have significant influence
on obtaining a balanced description of the spectrum in CVS methods.

\begin{figure}
	\includegraphics[width=0.95\columnwidth]{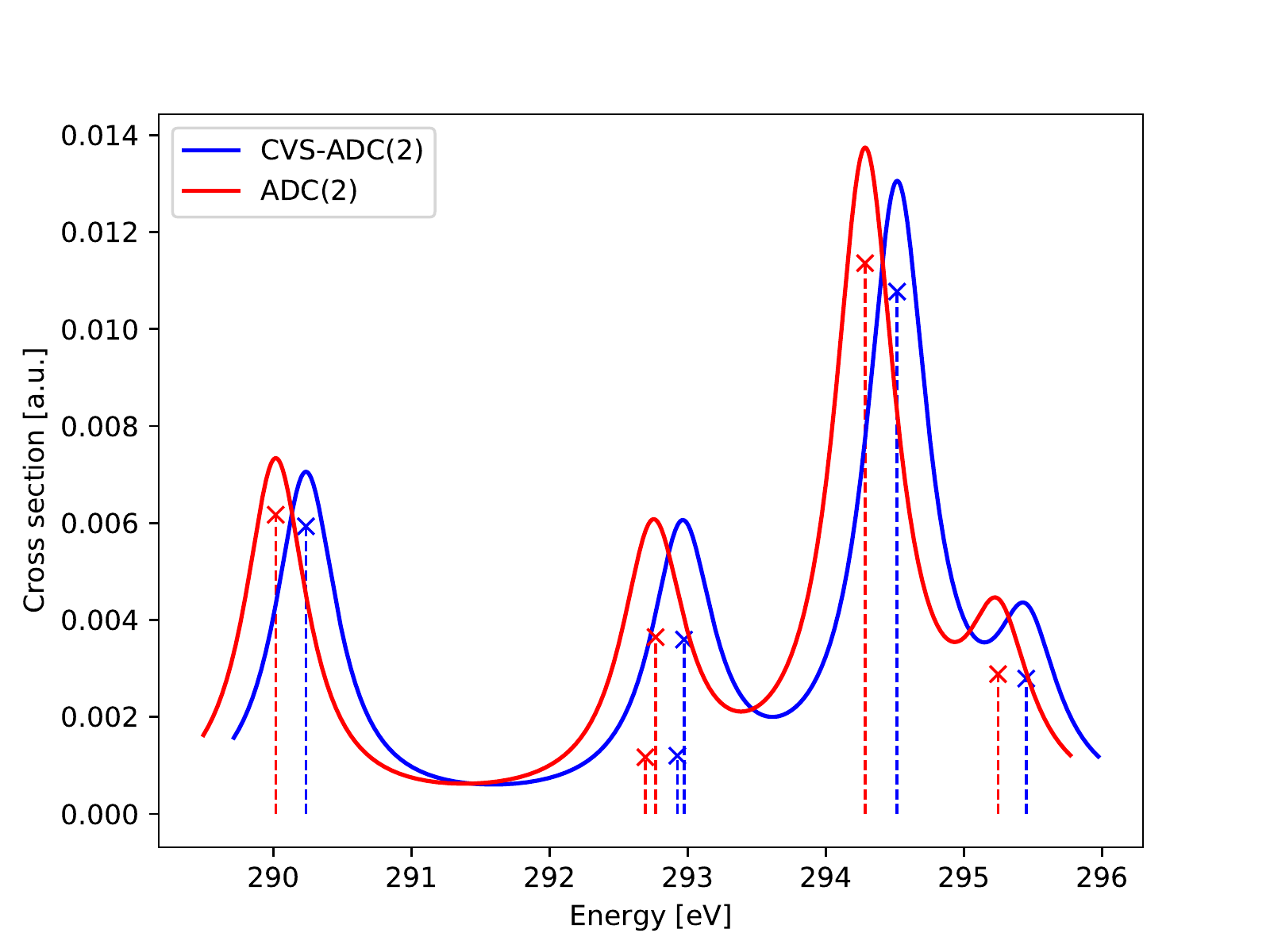}
	\caption{Comparison of core-excitation spectrum
		of the carbon edge of
		1,2-difluoroethane, as obtained at CVS-ADC(2) and ADC(2) levels
		using cc-pCVDZ for carbon and cc-pVDZ for all other elements.
		Peaks were broadened empirically using a Lorentzian line shape
		with width parameter \unit[0.01]{a.u.}.
		Between both methods the spectrum is primarily just shifted
		by about \unit[0.2]{eV}.
	}
	\label{fig:spectrum}
\end{figure}
Considering the dependency of the CVS error on the compound
and excitation,
errors for the oxygen edge are similar across all molecules
and transitions included here.
These span excitation energies from \unit[534.1]{eV} (HNO)
to \unit[540.7]{eV} (HFO).
In simulated excitation spectra this is visible by a scalar shift
of the complete spectrum when the CVS approximation is applied,
see Figure~\ref{fig:spectrum}.
For the carbon edge,
most errors are also found in a relatively narrow window,
save \ce{CO} and \ce{CS} as the two outliers.
For the latter two compounds the total CVS error is lower by
about \unit[0.05]{eV},
resulting in a larger CVS error spread for the carbon edge
compared to the oxygen edge.
Carbon transition energies span the range of
\unit[289.3]{eV} (\ce{CS}) to \unit[297.8]{eV} (\ce{CH2F2})
with the \ce{CO} resonance occurring at \unit[291.3]{eV}.

Using different basis sets, \ce{CO} and \ce{CS}
remain outliers, similarly when changing for
ADC(2)-x and ADC(3), although the magnitude of the deviations diminishes.
To a lesser extent \ce{HCN} is also a weak outlier
whereas \ce{HNC} (not included in the figure)
gives a total CVS error sitting between CO and HCN.
Additional calculations further show that
the deviating CVS errors
are not affected by restricting the CVS space
to only the carbon $1s$ or including also the nitrogen and sulphur $1s$.
We note that, e.g.~fluorine substitution,
which strongly shifts resonance energies,~\cite{fluoroethenes2013}
does not affect the CVS errors compared to related compounds.
This indicates that distortion of the electronic structure
near the core due to substitutions are not particularly influential on the CVS error, 
and we thus attribute the observed aberrations in \ce{CO} and \ce{CS}
to the distorted electronic structure around specifically triple-bond systems.
For practical purposes,
these distorted situations are unlikely to be a concern, and the
CVS error spread for carbon could thus be expected to be $\unit[\pm0.01]{eV}$,
i.e. of similar magnitude as the oxygen edge.
Furthermore, \ce{C2H2}, \ce{C2H4} and \ce{C2H6}
all have similar CVS errors as the other considered systems,
despite the fact that the $1s$ MOs are now delocalised over both carbon atoms.
This is encouraging, as delocalised $1s$ MOs can occur in larger systems.
Notice that such delocalisations are more of a concern in simulations,
as the real molecular system will always feature
vibrations or other disorder, which typically breaks the degeneracies in the $1s$.

\subsection{Comparison to previous analysis of the CVS error}
\label{sec:Comparision}

Previous studies have reported CVS errors of sizes varying from small negative
values up to one eV, in line with the order of magnitude we have obtained
in the previous Sections.
Within the ADC framework the work of \citet{cvsadc1985} reports
comparatively large errors (\unit[0.5--1.0]{eV})
of ADC(2) using second-order perturbation theory.
We account this to their work using a rather limited triple-zeta basis
similar to cc-pVTZ, where we also obtain values of this order.
Using damped response theory on ADC(2), \citet{adcrixs2017} have
estimated CVS errors for water of \unit[0.19--0.35]{eV}
on ADC(2) using a 6-311++G** basis,
which agrees with our results.

In the context of coupled cluster theory, Coriani and co-workers
have developed different CVS implementations.%
~\cite{ccxspec1,cvscc2015,cvseomccsd2019}
With linear-response coupled cluster
and a scheme similar to ours, they report CVS errors of
\unit[0.04]{eV} for neon, using an aug-cc-pCVTZ basis
with additional diffuse functions.~\cite{ccxspec1,cvscc2015}
Our results at the ADC(2) level is approximately \unit[0.1]{eV} lower,
around \unit[-0.06]{eV}.
This is within the spread of a few tenths of an eV we observed
in Section \ref{sec:method} comparing different ADC methods.

\citet{cumulantXray2019}
have recently developed two CVS schemes in the context of
linear-response density cumulant theory,
named CVS-ODC-12-a and CVS-ODC-12-b, and compared it to full ODC-12.
CVS-ODC-12-a is similar to the CVS scheme we employ,
neglecting both the cv-ov and the cv-cvcv couplings,
whereas CVS-ODC-12-b only neglects cv-ov.
For variant -a one would therefore expect similar discrepancies
as our total CVS errors, and for variant -b we instead expect
values similar to our singles-block CVS error.
Focusing on the first dipole-allowed feature of water using a 6-31G basis set,
the quoted CVS error is \unit[0.14]{eV} for variant -a
and \unit[-0.01]{eV} for variant -b.
Using ADC(2) we obtain a total CVS error of \unit[0.19]{eV}
and a singles-block error of \unit[-0.02]{eV},
whereas ADC(3) yields \unit[0.07]{eV} and \unit[-0.04]{eV}, respectively,
which is a reasonable agreement given the basis set.

\subsection{Comparison to experimental results}
\newcommand{\myspace}{\hspace{1.5em}}
\newcommand{\relativistic}{
\begin{minipage}{0.77\textwidth}
\footnotesize{\textsuperscript{a}%
Including scalar relativistic effects of 0.11, 0.21, 0.37, 0.61, 0.94 eV
for C, N, O, F, and Ne, respectively.}
\end{minipage}
}
\begin{table*}
\caption{
	Comparison of core-valence excitation energies
	obtained with ADC(2) and CVS-ADC(2) using a aug-cc-pCVTZ
	basis plus an additional diffuse $s$-function (exponent $10^{-11}$)
	on the probed element and cc-pVTZ on hydrogen.
	Energies are expressed in eV,\textsuperscript{a}
	$\Delta\text{Expt}$ denotes the error with respect to experimental values
	and $\Delta\text{CVS}$ the total CVS error.}
\label{table:expt_exci}
\begin{tabular}{ll@{\myspace}c@{\myspace}cc@{\myspace}cc@{\myspace}c}
	\hline \hline
	&&
		& \multicolumn{2}{c@{\myspace}}{ADC(2)}
		& \multicolumn{2}{c@{\myspace}}{CVS-ADC(2)}
		\\
\cline{4-5} 
\cline{6-7} 
	Molecule & Transition & Experiment
		& Energy & $\Delta\text{Expt}$
		& Energy & $\Delta\text{Expt}$ & $\Delta \text{CVS}$ \\
	\hline
\ce{CH4} & $1s \rightarrow 3p$      & 288.00~\cite{gaswater1993} & 290.57 & 2.67 & 290.63 &  2.74 & 0.07 \\[.3ex]
\ce{NH3} & $1s \rightarrow 3a(a_1)$ & 400.66~\cite{gaswater1993} & 402.68 & 2.22 & 402.72 &  2.26 & 0.04 \\
         & $1s \rightarrow 3p(e)$   & 402.33~\cite{gaswater1993} & 404.38 & 2.25 & 404.41 &  2.28 & 0.03 \\[.3ex]
\ce{H2O} & $1s \rightarrow 4a_1$    & 534.00~\cite{gaswater1993} & 535.36 & 1.72 & 535.36 &  1.73 & 0.01 \\
         & $1s \rightarrow 2b_2$    & 535.90~\cite{gaswater1993} & 537.20 & 1.67 & 537.20 &  1.67 & 0.00 \\[.3ex]
\ce{HF}  & $1s \rightarrow \sigma*$ & 687.40~\cite{fhxas1981}    & 687.86 & 1.07 & 687.84 &  1.04 &-0.02 \\[.3ex]
\ce{Ne}  & $1s \rightarrow 3p$      & 867.12~\cite{Coreno1999}   & 866.71 & 0.54 & 866.65 &  0.47 &-0.07 \\
	\hline \hline
\end{tabular}
\relativistic
\end{table*}
\begin{table*}
\caption{
	Comparison of ionisation potentials obtained with ADC(2) and CVS-ADC(2).
	Energies are in eV thoughout.\textsuperscript{a}
	For computational details see Table~\ref{table:expt_exci}.
}
\label{table:expt_ip}
\begin{tabular}{l@{\myspace}c@{\myspace}cc@{\myspace}cc@{\myspace}c}
	\hline \hline
	&
		& \multicolumn{2}{c@{\myspace}}{ADC(2)}
		& \multicolumn{2}{c@{\myspace}}{CVS-ADC(2)}
		\\
\cline{3-4} 
\cline{5-6} 
	Molecule &  Experiment
		& Energy & $\Delta\text{Expt}$
		& Energy & $\Delta\text{Expt}$ & $\Delta \text{CVS}$ \\
	\hline
	\ce{CH4} & 290.76~\cite{ch4ip1979}    & 292.08 &  1.43 & 292.15 &  1.49 &  0.07 \\
	\ce{NH3} & 405.52~\cite{nh3ip1976}    & 405.57 &  0.26 & 405.60 &  0.29 &  0.03 \\
	\ce{H2O} & 539.90~\cite{gaswater1993} & 538.34 & -1.19 & 538.33 & -1.20 & -0.01 \\
	\ce{HF}  & 694.10~\cite{fhip1980}     & 691.36 & -2.13 & 691.36 & -2.17 & -0.04 \\
	\ce{Ne}  & 870.09~\cite{neonip1080}   & 866.39 & -2.76 & 866.33 & -2.82 & -0.06 \\
	\hline \hline
\end{tabular}
\relativistic
\end{table*}

A comparison of the performance of CVS-ADC(2)
and full ADC(2) with respect to experimental
data is reported in Tables~\ref{table:expt_exci} and~\ref{table:expt_ip}
for excitations to bound states and for ionisation potentials, respectively.
For the excitations to bound states
both CVS-ADC(2) and ADC(2) show a decreasing error
along the elements of the second period.
Neon deviates a little from this trend due to the inability
of the employed basis to describe the $1s\rightarrow3p$ transition properly.
Since transition energies are overestimated for all systems
and the CVS error is mostly positive, ADC(2) agrees a little better
with experiment than CVS-ADC(2).
For estimating the ionisation potential
both ADC(2) models overestimate experiment for low $Z$ and
underestimate it for high $Z$, again with CVS relaxation
improving results a little.
The larger discrepancy observed for ionisation is unsurprising,
since it has been previously noted
that fourth-order ADC methods are required for
a good description of ionisation potentials.~\cite{adcip2003}
Clearly for both excitation and ionisation the
CVS error is negligible compared to the total error with respect to experiment, and the difference in the CVS error between the two processes is smaller still.

\section{Conclusions and outlook}
A simple Rayleigh-Quotient-based
scheme for expanding eigenstates obtained with the core-valence separation
(CVS) scheme to full space eigenstates for the ADC($n$) hierarchy has been
developed and implemented
within the framework of the adcc python module.~\cite{adcc2019}
This scheme is general and could be applied to relaxing
the CVS approximation in other contexts such as coupled-cluster
approaches and linear-response time-dependent density-functional theory.
Using this approach,
the error imposed by the CVS approximation has been investigated
for $K$-edge absorption spectrum calculations,
yielding discrepancies in energy of \unit[-0.4]{eV} to \unit[0.7]{eV},
with exact value depending on
the ADC level, basis set, and element in consideration.

The CVS error can to leading order be identified with
two kinds of coupling terms, which are neglected in the CVS-ADC matrix,
namely (1) cv-ov terms coupling single core- and single valence-excitations
and (2) cv-cvcv terms, which describe the interaction
of single and double core-excitations.
Neglecting cv-ov and cv-cvcv has counteracting effects,
since the cv-ov terms were identified to increase transition energies
and the cv-cvcv to decrease them.
Including cv-cvcv terms only, as is done in some CVS implementations,
can thus yield a larger CVS error for the $1s$-core transitions
discussed in this article.
Large cv-ov and cv-cvcv couplings were further identified
to be a consequence of a balanced description of core and valence region.
Suitable basis sets for a description of core-excitations at CVS level
therefore give rise to large values
of equal magnitude for these couplings,
resulting in a small CVS error albeit a good description of the physics.
This has been shown to be the case for basis sets with
additional tight functions of a core-polarising nature.

The CVS error is typically
larger for ADC(2) than for ADC(2)-x or ADC(3), and focusing on the carbon and
oxygen edge of a set of representative compounds
we report error spreads of maximally \unit[0.02]{eV}
for bound excitations and ionisations, separately,
with absolute errors being around \unit[0.2]{eV}
for suitably core-polarised basis sets.
Errors for the transition to extremely diffuse states
are systematically about \unit[0.02--0.03]{eV} lower.
Larger deviations from the typical CVS error on the carbon edge
are only noted for \ce{CO} and \ce{CS}.
This has been attributed to the strong changes of
electronic structure surrounding triple bounds.
For most practical considerations the CVS approximation
only yields a small error,
which amounts to a mere
a scalar shift of the full spectrum.
The size of this shift is at least an order of magnitude
smaller than the error with respect to experiment,
making the CVS approximation a justified tool for the calculation
of core-excitation spectra.



\section{Acknowledgments}
The authors thank Andreas Dreuw and Jochen Schirmer
for stimulating discussions.
T.F. acknowledges financial support from the Swedish Research Council
(Grant No.~2017-00356).
This project has received funding from the European Research Council
(ERC) under the European Union's Horizon 2020 research and innovation
program (grant agreement No~810367).

\section*{AIP publishing data sharing policy}
\noindent
The data that support the findings of this study are openly available on
github (\url{https://github.com/mfherbst/cvs-relaxation-scripts})
and have been deposited on Zenodo with DOI 10.5281/zenodo.3824506.

\section*{Appendix: Leading-order perturbative correction to the CVS excitation energies}
The full ADC matrix can be written as $2\times2$ blocks
\begin{equation}
	\mat{M} = \left(\begin{array}{cc}
		\mat{M}^\text{core} & \mat{C} \\
		\mat{C}^\dagger & \mat{M}^\text{rest}
		\end{array}\right),
\end{equation}
where $\mat{M}^\text{core}$ consists
of the blocks kept in our CVS scheme,
i.e.~the ones involving only cv singly and cvov doubly excited configurations,
$\mat{M}^\text{rest}$ collects all others (ov, cvcv, ovov) and
$\mat{C}$ denotes the coupling between both classes.
In this partitioning \mbox{$\mat{M}^\text{core} = \mat{m} + \mat{D}$},
where $\mat{m}$ is the CVS matrix as defined in Section \ref{sec:relxation}.
Notice that both $\mat{D}$ and $\mat{C}$ are zero
if one neglects the coupling of core and valence orbitals
via the Coulomb kernel~\cite{WenzelPhd}
and thus represents the terms missed by a CVS treatment.
To apply perturbation theory we split $\mat{M}$ as
\begin{equation}
	\mat{M} =
	\underbrace{
	\left(\begin{array}{cc}
		\mat{m} & 0 \\
		0 & \mat{M}^\text{rest}
		\end{array}\right)
	}_{=\mat{M}^{(0)}}
	+ \underbrace{
	\left(\begin{array}{cc}
		\mat{D} & \mat{C} \\
		\mat{C}^\dagger & 0
		\end{array}\right)
	}_{=\mat{M}^{(1)}}.
\end{equation}
As reference states we take the CVS vectors $\vec{x}_i$
from Equation \eqref{eqn:cvs_diagonalisation},
which are eigenstates of $\mat{M}^{(0)}$ with eigenvalue $\omega_i$
when extended by rows of zeros.
In the following we restrict ourselves
to the leading-order contribution in the perturbation $\mat{M}^{(1)}$,
which amounts to keeping only the ADC($1$) terms in $\mat{C}$
and neglecting $\mat{D}$ completely,
where the lowest-order terms are at ADC($2$) level.
With this simplification the correction for the excitation energy of $\vec{x}_i$
is to second order
\begin{equation}
	\sum_I \frac{\left|\vec{x}_i^\dagger \mat{C} \vec{y}_I\right|^2}{\omega_i - E_I}.
\end{equation}
In this expression $I$ runs over eigenstates
$\vec{y}_I$ of $\mat{M}^\text{rest}$ with corresponding eigenvalues $E_I$.
Since $E_I$ is to leading order given by orbital energy differences,
$E_I < \omega_i$ for $I$ being an ov-dominated excitation
and $E_I > \omega_i$ for $I$ being a cvcv-dominated excitation.
For ovov-dominated excitations the coupling term in the numerator
becomes zero at ADC($1$) level,
such that these terms do not contribute to lowest order.
Overall including the neglected cv-ov coupling therefore
pushes the energy of the core-excitations up,
the cv-cvcv coupling pushes the energy down and
the cv-ovov coupling has little effect.
\bibliography{thomas}
\end{document}